\documentclass[preprint,final,1p,article]{elsarticle}
\usepackage{lineno}
\modulolinenumbers[2]
\journal{Nuclear Physics A}
\usepackage{amssymb}
\usepackage{graphicx}
\usepackage{color}
\usepackage{amsthm}
\usepackage{amsmath}
\usepackage{amssymb}
\usepackage[colorlinks,citecolor=blue,linkcolor=red,anchorcolor=blue,filecolor=blue,urlcolor=blue]{hyperref}
\begin{document}
\begin{frontmatter}
\title{ Competing decay modes and stability analysis of superheavy nuclei with Z = 120 using relativistic mean-field approach}
\author[1]{Nishu Jain}
\ead{nishujain1003@gmail.com}
\author[2]{M. Bhuyan}
\ead{bunuphy@yahoo.com}
\author[3]{P. Mohr}
\ead{mohr@atomki.hu}
\author[1]{Raj Kumar}
\ead{rajkumar@thapar.edu}
\address[1]{Department of Physics and Materials Science, Thapar Institute of Engineering and Technology, Patiala 147004, India}
\address[2]{Institute of Physics, Sachivalya Marg, Bhubaneswar-751005, Odisha, India}
\address[3]{Institute of Nuclear Research (ATOMKI), H-4001 Debrecen, Hungary}

\begin{abstract}
\noindent
We systematically study the competition between $\alpha$-decay and spontaneous fission in $even–even$ superheavy nuclei with (Z=120) and $256 \leq A \leq 304$ within the preformed cluster-decay model using microscopic inputs from relativistic mean-field calculations with the NL3 parameter set. The $\alpha$-decay half-lives are obtained from WKB barrier penetration with empirically determined preformation factors, self-consistent $Q_\alpha$ values from RMF, and nuclear interaction potentials constructed using both M3Y and relativistic R3Y nucleon–nucleon forces, and are benchmarked against standard semi-empirical formulas. Our results predict reduced spontaneous fission probabilities and extended $\alpha$-decay chains toward the fermium region for isotopes with $296 \leq A \leq 304$, with enhanced stability reflected in maxima of $\log_{10} T_{1/2}$ around neutron numbers $N \approx 166–182$. In particular, the nuclei $^{296,298,300,302,304}$120 are identified as the most favorable candidates for survival against fission, demonstrating the crucial role of shell effects, deformation, and pairing correlations and providing quantitative guidance for future experimental searches of $Z=120$ nuclei.
\end{abstract}
\begin{keyword}
\texttt{Superheavy Nuclei, Alpha Decay, Spontaneous Fission, Relativistic Mean Field, Preformed Cluster Decay Model, Shell Effects, Nuclear Stability}
\end{keyword}
\end{frontmatter}
\section{Introduction}
\label{intro}
Exploring new frontiers in low-energy nuclear physics, particularly in the study of superheavy elements (SHNs), remains an exciting endeavor for the scientific community. This region continues to attract both experimental and theoretical interest because of its unique structural features and the ongoing effort to extend the nuclear landscape toward higher $Z$ values \cite{mois25,ogan13}. The systematic relationship between $\alpha$-decay half-lives and the kinetic energy (K.E.) of emitted alpha particles was first identified by Geiger and Nuttall in 1911 \cite{geig11}. Alpha decay remains one of the most important probes for establishing the existence and stability of heavy and superheavy nuclei. The quantum shell effects stabilise superheavy nuclei (SHNs) by creating significant energy gaps at magic or semi-magic nucleon numbers, which counteract the strong Coulomb repulsion and weak binding forces that dominate in other regions of the nuclear chart. The quantum tunnelling process has been convincingly explained as the cause of alpha decay \cite{gamo28,cond28,moll97}. Experimentally, $\alpha$-decay chains from an unknown to a known region of the chart of nuclei are utilised to discover and identify the new elements. Until now, many theoretical studies on the $\alpha$-decay have been performed using a variety of models and methods, including cluster models (CM) \cite{Buck94}, density-dependent cluster models (DDCM) \cite{Ni10}, the analytical superasymmetric (ASAF) model \cite{Poen95}, the macroscopic-microscopic model (MMM) \cite{moll97,Peng08}, the Viola-Seaborg formula \cite{Sobi89}, the Skyrme-Hartree-Fock (SHF) model \cite{Gori01}, microscopic methods \cite{Pogg69}, the generalized liquid drop model (GLDM) \cite{Roye00}, etc. Alongside these theoretical models, various empirical formulae have been devised to accurately reproduce the experimental data of cluster decays. However, it is important to note that many of these models depend on binding energies obtained from experimental data \cite{wang12} and the Finite Range Droplet Model (FRDM) \cite{moll97,moll19}, whose reliability becomes increasingly uncertain in the unexplored and extreme regions of the nuclear chart. The nuclear potential used in the various theoretical models can also be calculated with the relativistic mean-field (RMF) theory \cite{Vret05,Meng06}. The strength of the RMF framework lies in its robust theoretical foundation, rooted in relativistic quantum field theory, which enables reliable predictions across the nuclear chart, especially in regions where experimental data are scarce or nonexistent. This makes RMF particularly valuable for exploring the structure and decay properties of superheavy nuclei, where experimental access is limited and theoretical input becomes essential. Its ability to naturally incorporate key nuclear features such as spin-orbit coupling, shell effects, and nuclear deformation further enhances its predictive power in these extreme regimes. Recently, Mohr \textit{et al.} \cite{mohr25} provided a comprehensive evaluation of $\alpha$-nucleus potentials, highlighting the Atomki-V2 potential as a reliable model for predicting $\alpha$-induced reaction rates on intermediate and heavy nuclei. Furthermore, the recent investigation of the superheavy nuclei (SHN), the island of stability has become one of the hottest topics in nuclear physics \cite{Ogan10} and Refs. therein. The stability of the newly synthesised SHN \cite{xu08} is affected by both $\alpha$-decay and spontaneous fission, and precise alpha-decay predictions are crucial for future experiments and identification.\\ \\
Identifying the structural properties of nuclei in the superheavy mass region is a tricky problem, but it is more useful to comprehend the concept of the ``island of stability" beyond the doubly magic nucleus. More recent microscopic calculations have indicated a range of different stable regions, including $Z$ = 114, $N$ = 184 \cite{Rutz97}; $Z$ = 120, $N$ = 172 or 184 \cite{Gupt97,Patr99}; and $Z$ = 124 or 126, $N$ = 184 \cite{Cwio05}. The magic proton number at $Z$ = 120, 132, 138, and the magic neutron number at $N$ = 172, 184, 198, 228, 238, 258 are also predicted in the framework of relativistic continuum Hartree-Bogoliubov theory \cite{Zhan05}. Taninah \textit{et al.} \cite{Tani20} conducted a comprehensive study of the bulk nuclear and fission properties of even-even actinides and superheavy nuclei with $Z$ = 90-120. Possible shell closure at  $Z$ = 120, $N$ = 184 or 258 is predicted using the covariant energy density formalism for the DD-PC1, DD-ME2, NL3 and PC-PK1 parameter sets. Superheavy nuclei, characterised by their immense atomic mass, typically progress through consecutive $\alpha$-decay chains, culminating in spontaneous fission. Identifying newly synthesised superheavy nuclei hinges upon analysing the decay products of these $\alpha$-decay chains. However, cluster radioactivity is another intriguing decay phenomenon that mirrors the mechanism of $\alpha$-decay. Pioneering research by Poenaru \textit{et al.} \cite{poen10}, and this work suggests cluster decay could challenge $\alpha$ decay and spontaneous fission in select isotopes of superheavy nuclei. The former relegates the concept of preformation; that is, it is assumed that clusters are formed as the parent nucleus separates and deforms while breaking through the barrier of confining interactions. The preformed cluster-decay model (PCM) \cite{Gupt88,kuma12,Mali89,gupt99}, which is based on the well-known quantum mechanical fragmentation theory (QMFT) \cite{maru74,fink74} is one of the models that incorporate this concept. In the PCM, an alpha particle is pre-born within the parent nucleus and then tunnels across the potential barrier formed by the combination of the nuclear and Coulomb potentials \cite{gamo28,cond28}. The Coulomb potential is straightforward to calculate; however, the nuclear potential can be calculated using phenomenological \cite{horn12} and microscopic methods \cite{schu16}. As a result, the selection of nuclear potential is crucial for obtaining a thorough understanding of the decay processes. Apart from the fundamental concepts \cite{ekst13}, the recently developed R3Y nucleon-nucleon potential \cite{sing12}, which is comparable to the phenomenological M3Y \cite{Satc79}, is deduced from the relativistic mean-field (RMF) Lagrangian. The R3Y NN potential is used in the present work to investigate the emission using the NL3 parameter set, which has been satisfactorily used in the analysis of various ground and excited state properties \cite{Bisw20}. Nuclear property estimations, such as half-lives of nuclei that are noted on the stability of islands, are another fascinating aspect of $\alpha$-decay research. The Q-values play an important role in the calculation of alpha-decay half-lives. Various empirical and theoretical relations can be used to calculate the Q-values \cite{Moll95,Wang14}. Typically, SHN has a relatively short $\alpha$-decay half-life. Previously, similar calculations have been performed for various isotopic chains where experimental data are available \cite{Bisw21,Jain22,maje23}. The obtained theoretical results show good agreement with the measured values, which provides confidence in extending the present theoretical framework to predict the decay properties of nuclei in the superheavy region where experimental information is still lacking.\\\\
In the present work, we have investigated the alpha decay properties of the nuclei with $100\le Z \le 120$ and $158\le N \le 184$ neutron numbers to determine the possible decay modes and the stability of the nuclei considered using the RMF formalism. The WKB approximation \cite{Went26,Kram26,Bril26} is used to determine the penetration probability by presuming $\alpha$-particle tunnelling through the potential barrier between the $\alpha$-cluster and the daughter nucleus. The decay constant $\alpha$ is generally calculated as the product of three quantities: particle preformation probability $P_0$, assault frequency $\nu_0$, and barrier penetration probability $P$. The major factors determining the half-lives are the height, position, and width of potential barriers. In the present work, the input of PCM includes the preformation probability, calculated from the analytical formula \cite{maje23}, and the penetration probability (P), using the WKB approximation \cite{Went26,Kram26,Bril26}. The influence of the relative separation distance between two fragments or clusters in the PCM is expressed through the neck length $\Delta R$, which accounts for the effect of neck formation and identifies the initial turning point of barrier penetration. In this work, all inputs like binding energies, densities, and interaction potentials are derived self-consistently from the same RMF framework. This uniform treatment links nuclear structure directly to decay dynamics without empirical tuning and offers a coherent microscopic basis for predicting $\alpha$-decay and fission behaviour in yet-unobserved nuclei. The calculated $Q$-values are compared with the available experimental data \cite{wang12} and subsequently employed in the evaluation of decay half-lives. The half-life calculations assume a complete $\alpha$-decay branching ratio (100$\%$) and are performed without any additional parameter fitting and/or empirical adjustment, thereby ensuring a fully theoretical prediction. For further validation, the present results are also compared with those obtained using the M3Y interaction and other well-known semi-empirical formulas, such as the universal decay law (UDL) and the Horoi formula \cite{isma24}. In addition to $\alpha$-decay, the spontaneous fission (SF) half-lives are also evaluated using the semi-empirical formulation of Xu \textit{et al.} \cite{xu08}, to examine the competing decay modes. The relative probability of the two decay channels is quantified through the branching ratio $b = T_{\mathrm{SF}}/T_{\alpha}$, which identifies the dominant decay mode for each isotope. \\ \\
The following is a description of the structure of this paper: The ensuing section outlines the theoretical framework employed to calculate the half-lives of $\alpha$ decay. Sec. \ref{sec:2} introduces the theoretical framework. Sec. \ref{results} contains the findings and related discussions. Sec. \ref{result} contains the overall conclusion of the paper.
\section{Theoretical framework} \label{sec:2}
\noindent
The Quantum hydro-dynamics (QHD) mean field method has traditionally been used to characterise the nuclear structure and infinite nuclear matter properties \cite{Bhuy15,lala99}. The nucleus is considered to be a coupled system of nucleons (neutron and proton) interacting via the interchange of mesons and photons in the relativistic mean field theory \cite{Vret05,Meng06,Ring96,Rein89}. The contribution from the meson fields is described as mean fields or point-like interactions between the nucleons \cite{Burv02}, and non-linear coupling components \cite{Broc92} or density-dependent coupling constants \cite{Niks08} are included to reflect the accurate saturation features of infinite nuclear matter. For a nucleon-meson many-body system \cite{Vret05,Meng06,Bhuy15,lala99,Ring96,Rein89,Bhuy18,Bhuy20}, the relativistic Lagrangian density is:
\begin{eqnarray}
{\cal L}&=&\overline\psi_i\left\{i\gamma^{\mu}\partial_{\mu}-M\right\}\psi_i+\frac{1}{2}\partial^{\mu}\sigma\partial_{\mu}\sigma\nonumber\\
&&-\frac{1}{2}m_{\sigma}^{2}\sigma^2-\frac{1}{3}g_2\sigma^3-\frac{1}{4}g_3\sigma^4-g_s\overline{\psi}_i\psi_i\sigma\nonumber\\
&&-\frac{1}{4}\Omega^{\mu\nu}\Omega_{\mu\nu} +\frac{1}{2}m^2_\omega V^\mu V_\mu-g_\omega\overline{\psi}_i\gamma^\mu\psi_i V_\mu\nonumber\\
&&-\frac{1}{4}\vec B^{\mu\nu}.\vec B_{\mu\nu}+\frac{1}{2}m^2_\rho\vec R^\mu.\vec R_\mu-g_\rho\overline{\psi}_i\gamma^\mu\vec{\tau}\psi_i.\vec R^\mu\nonumber\\
&&-\frac{1}{4}F^{\mu\nu}F_{\mu\nu}-e\overline{\psi}_i\gamma^\mu(\frac{1-\tau_{3i}}{2})\psi_i A_\mu. \label{1}
\end{eqnarray}
The medium-range attraction and short-range repulsion between nucleons are represented by the scalar $\sigma$ meson and the vector meson $V_{\mu}$, respectively. The isospin of nuclei effects are described by isovector-vector meson $\vec R^\mu$. Their masses are $m_\sigma$, $m_\omega$ and $m_\rho$ respectively  with the coupling constants $g_s$, $g_\omega$ and $g_\rho$. The Dirac spinor, isospin and its third component are represented as $\psi_i$, $\tau$ and $\tau_3$ respectively. The coupling constants for the non-linear terms are $g_2$, $g_3$ and $\frac{e^2}{4\pi}$. The mass of the nucleon is $M$, and the electromagnetic field is $A_\mu$. Because of pseudo-scalar character \cite{Ring96}, the contribution of $\pi$-meson is negligible. The Dirac equation is deduced from equation (\ref{1}) using the classical variation principle,
\begin{equation}
    [-i\alpha.\nabla+\beta(M^*+g_\sigma\sigma)+g_\omega\omega+g_\rho\tau_3\rho_3]\psi_i=\epsilon_i\psi_i\label{2}
\end{equation}
to get the Klein-Gordon equations and nuclear spinors
\begin{eqnarray}
      (-\nabla^2+m^2_\sigma)\sigma(r)&=&-g_\sigma\rho_s(r)-g_2\sigma^2(r)-g_3\sigma^3(r),\nonumber\\
      (-\nabla^2+m^2_\omega)V(r)&=&g_\omega\rho(r),\nonumber\\
   (-\nabla^2+m^2_\rho)\rho(r)&=&g_\rho\rho_3(r). \label{3}
\end{eqnarray}
The numerical solution is then performed by using an iterative approach with the NL3$^*$ parameter adjusted in a self-consistent manner \cite{lala09}. The scalar, $\sigma$- and vector $(\omega,\rho)$-fields are defined in terms of the NN potential as for a single-meson exchange in a static baryonic medium. 
 \begin{eqnarray}
   V_\sigma= -\frac{g^2_\sigma}{4\pi}\frac{e^{-m_\sigma r}}{r}+\frac{g^2_2}{4\pi}re^{-2m_\sigma r}+\frac{g^2_3}{4\pi}\frac{e^{-3m_\sigma r}}{r},\nonumber\\
   V_\omega(r)=\frac{g^2_\omega}{4\pi}\frac{e^{-m_\omega r}}{r},\hspace{5mm} V_\rho(r)=+\frac{g^2_\rho}{4\pi}\frac{e^{-m_\rho r}}{r}. \label{4}
 \end{eqnarray}
 The effect of $\delta$-meson is incorporated in the $\rho$-field \cite{Ring96} and therefore, $V_{\delta}$ is insignificant. The RMF-based R3Y NN effective interactions plus a single-nucleon exchange \cite{Gupt65} are obtained by adding the starting NN interactions in equations (\ref{4})
    \begin{eqnarray}
    V_{\mbox{eff}}^{R3Y}(r)&=&\frac{g^2_\omega}{4\pi}\frac{e^{-m_\omega r}}{r}+\frac{g^2_\rho}{4\pi}\frac{e^{-m_\rho r}}{r}-\frac{g^2_\sigma}{4\pi}\frac{e^{-m_\sigma r}}{r}\nonumber\\
   &&+\frac{g^2_2}{4\pi}re^{-2m_\sigma r}+\frac{g^2_3}{4\pi}\frac{e^{-3m_\sigma r}}{r}+J_{00}(E)\delta(s), \label{5}
 \end{eqnarray}
 where $J_{00}(E)=-276(1-0.005 E_\alpha/A_{\alpha})$ MeV fm$^3$. The mass of the $\alpha$-particle is represented by $A_{\alpha}$ and the energy $E_{\alpha}$ determined in the center-of-mass of the decay fragments ($\alpha$-daughters) systems is equal to the energy released during the $\alpha$-decay process ($Q_\alpha$-value). Unlike the energies necessary in high-energy scattering, $J_{00}(E)$ could be employed in its estimated form and is independent of $Q$-value \cite{Basu03}. \\
 The form of relativistic R3Y NN potential is similar to the frequently used M3Y potential, which includes Michigan-3-Yukawa having 0.25 fm medium-range attractive part, 0.4 fm short-range repulsive part and 1.414 fm long-range tail of one-pion exchange potential (OPEP). This result from the fitting of $G-$matrix elements in an oscillator basis depends on Reid-Elliott soft-core NN-interaction \cite{Satc79}. The M3Y plus exchange term has the following equation:
 \begin{equation}
    V_{\mbox{eff}}^{M3Y}(r)=7999\frac{e^{-4r}}{4r}-2134\frac{e^{-2.5r}}{2.5r}+J_{00}(E)\delta(r), \label{6}
\end{equation}
Here, the unit of range is in FM, and the strength is in MeV. The nuclear interaction potential $V_n (R)$ is determined using the M3Y and R3Y nucleon-nucleon potentials in the double folding method \cite{Satc79}. The nuclear potential is expressed as 
\begin{equation}
    V_n(R)=\int\rho_{\alpha}(\vec r_{\alpha})\rho_{d}(\vec r_{d})V_{eff}(|\vec r_{\alpha}-\vec r_{d}+\vec R|\equiv r)d^3 r_{\alpha}d^3 r_{d}. \label{7}
\end{equation}
The nuclear matter density distributions of the $\alpha$-particle ($\alpha$) and the daughter nucleus (d) are shown by $\rho_{\alpha}$ and $\rho_{d}$. The nuclear potential $V_n(R)$ is combined to the Coulomb potential $V_C(R)$ $=\frac{Z_{c}Z_{d}}{R}e^2$ and is given as
\begin{eqnarray}
V(R)= V_n (R)+V_C(R).
\label{8}
\end{eqnarray}
This potential is used in the Preformed cluster-decay model (PCM) to calculate the WKB penetration probability.
The alpha decay half-life and decay constant are calculated in the PCM, as follows \cite{Mali89,Gupt94}:
\begin{equation}
    T^{\alpha}_{1/2}=\frac{\ln2}{\lambda},   \hspace{0.5cm}  \lambda= \nu_{0} P_0 P. \label{12}
\end{equation} 
The decay constant $\lambda$ denotes the probability per unit time for each nucleus to decay. Clusters are thought to be pre-born with a particular preformation $P_0$ within the parent nucleus, hits the potential barrier with an impinging frequency $\nu_0$, given as
\begin{equation}
    \nu_{0}=\frac{\mbox{ velocity }}{R_0}=\frac{\sqrt{2E_{\alpha}/\mu}}{R_0} \label{10}.
\end{equation}
Afterwards, tunnels with a probability $P$ and $R_0$ denote the radius of the parent nucleus. $E_\alpha$ stands for the kinetic energy (K.E.) of the emitted $\alpha$ particle. A positive Q-value is an essential condition for the energetically preferred spontaneous emission of $\alpha$-particle. This is the maximal amount of energy that can be used in the $\alpha$-decay process. 
 \begin{equation}
     Q=BE_p-(BE_d+BE_\alpha), \label{11}
 \end{equation}
The ground state binding energies of the parent, daughter nuclei and $\alpha$-cluster are represented as $BE_p$, $BE_d$ and $BE_c$, respectively. The Q-values are conserved between two fragments, i.e., $\alpha$-particle $E_\alpha=\frac{A_d}{A}Q$ and $E_d=Q-E_\alpha$ for the recoil energy of the daughter nuclei since $Q=E_\alpha+E_d$. 
The initial turning point $R_a$ depicts the penetration path of the decaying CN and is defined as:
\begin{equation}
     R=R_a=R_t+\Delta R.\label{15}
\end{equation}
The neck length parameter is defined as the relative separation distance $\Delta R$ between two outgoing nuclei, which takes into account the neck formation effects between them. The neck is optimised to determine the experimental measurements of the half-lives. It's worth noting that the reaction Q-value has a significant effect on the neck-length selection. As a result, the potential at the initial turning point $V(R_a)$ must be greater than the Q-value. Starting at the 1$^{st}$ classical turning point $R=R_a$, the $\alpha$-particle tunnels through the interaction potential $V(R)$ until it reaches the 2$^{nd}$ turning point $R=R_b$, where the equivalent potential $V(R_b)=Q$ for ground state decays. Furthermore, $V(R_a)=Q+E_i$ \cite{Mali89}, where $E_i$ is the energy with which the $\alpha$-particle or daughter nucleus decays into an excited state. The structure of the parent nucleus alters as instability sets in, causing the separation of the $\alpha$-particle and neck formation. \\ 
Furthermore, within the PCM framework, it is assumed that both daughter nuclei and clusters originate from the ground state (g.s.) with a certain preformation probability, denoted as $P_0$. This parameter incorporates the structural characteristics of the decaying parent nucleus. However, determining the precise value of $P_0$ in a microscopic context can be challenging, given the complexities of the nuclear many-body problem. Nonetheless, empirical evidence suggests that $P_0$ in alpha decay scenarios tends to be of the order of magnitude below unity. Therefore, to assess the preformation probability ($P_0$) for alpha decay is:\\
\begin{equation}
    logP_0 = -\frac{aA_\alpha\eta_A}{r_B} - Z_\alpha\eta_Z + bQ + c. 
    \label{18}
\end{equation}
Here, the values of constant parameters a, b, and c are 11.98, 0.037, and 1.52, respectively.
{The asymmetry parameters ($\eta_A$) and ($\eta_Z$), along with the barrier radius ($r_B$), are defined as follows: $\eta_A = \frac{A_d - A_c}{A_d + A_c},$
$\eta_Z = \frac{Z_d - Z_c}{Z_d + Z_c}$,
$r_B = 1.2\,{\rm{fm}} \times (A_c^{1/3} + A_d^{1/3})$.} 
More details about preformation probability are found in Ref. \cite{maje23}.\\ 
\subsection{Semi-empirical Formulae}
Several empirical formulas, such as the universal decay law (UDL) and the Horoi formula, have been developed to reproduce experimental data for alpha decays. \\ \\
\textbf{Universal decay law (UDL) formula}\\
Qi \textit{et al.} \cite{qi09} formulated a linear universal decay law (UDL) based on the R-matrix theory, delineating the microscopic mechanism of charged-particle emission, which applies to both $\alpha$ decays.  In our investigation, we employ the UDL formula, expressed as follows:\\
\begin{eqnarray}
\log_{10} T_{1/2}^{UDL} &= & aZ_\alpha Z_d \sqrt{\frac{A}{Q_\alpha}} + b\sqrt{\mu Z_\alpha Z_d(A_d^{1/3} + A_\alpha^{1/3}}) + c \nonumber \\
&& = a\chi' + b\rho' + c. 
\label{qi-alpha}
\end{eqnarray}
%
{Here, reduced mass is defined as $\mu = A_dA_\alpha/(A_d + A_\alpha$), where $A$ denotes the mass number of the parent nucleus, $A_d$ represents the mass number of the daughter nucleus, and $A_\alpha$ corresponds to the mass number of the emitted $\alpha$-particle.}
The energy released from the alpha decay, denoted as  $Q_\alpha$, is computed using the mass excess of the nuclei involved \cite{wang21}. In the preceding equation, the coefficients for the UDL formula are designated as follows: $a$ = 0.3949, $b$ = -0.3693, and $c$ = -23.7615 \cite{isma24}. \\ \\
\textbf{HOROI formula}\\
In 2004, Horoi \textit{et al.} \cite{horo04} introduced a semi-empirical formula (HF) for cluster decay, which has also been extensively applied to $\alpha$-decay. Saxena \textit{et al.} \cite{shar21} enhanced this formula by incorporating asymmetry dependence (MHF), effectively describing $\alpha$-decay chains while considering the competition between $\alpha$-decay and spontaneous fission. The development of a new modified Horoi formula (NMHF) follows as:\\
\begin{eqnarray}
   \log_{10} T_{1/2}^{NMHF}=(a\sqrt{\mu+b})[{(Z_\alpha Z_d)}^{0.6}Q_\alpha^{-1/2}-7]+\nonumber\\
   (c\sqrt{\mu}+d)+eI+fI^2+gl(l+1)\nonumber\\
\end{eqnarray}
In the above equations, $\mu$ is the reduced mass, which is given by $A_dA_\alpha$/($A_d$ + $A_\alpha$) where $A_d$ and $A_\alpha$ are the mass numbers of the daughter nucleus and $\alpha$-particle, respectively. Likewise, $Z_d$ and $Z_\alpha$ represent the atomic number of the daughter nucleus and $\alpha$-particle, respectively, and $Q_\alpha$ (in MeV) is the energy released in ground-state to ground-state $\alpha$-decay. I =(N-Z)/A is the nuclear isospin asymmetry, which has been demonstrated to play an essential role in the determination of the half-life of $\alpha$-decay. More details can be found in Ref. \cite{isma24}.\\ \\
\textbf{Spontaneous fission half-lives}\\
Xu \textit{et al.} \cite{xu08} proposed a semi-empirical method for estimating the half-lives of spontaneous fission (SF) based on the parabolic potential approximation. This approach incorporates nuclear structure effects and assumes an inverted parabolic shape for the potential barrier during fission. The formula reads as follows:\\
\begin{eqnarray}
    T_{1/2}=exp\Bigl\{2\pi\Bigl[C_0 + C_1A_P + C_2Z_P^2 + C_3Z_P^4 +\nonumber\\
    C_4(N_P-Z_P)^2 - \Bigl(0.13323\frac{Z_P^2}{A_P^{1/3}}-11.64\Bigr)\Bigr]\Bigr\}
\end{eqnarray}
where, the values of constants are $C_0$ = -195.09227, $C_1$ = 3.10156, $C_2$ = -0.04386, $C_3$ = 1.4030 × $10^{-6}$, and $C_4$ = -0.03199.
\section{Results and Discussions}
\label{results} 
\begin{figure}[!h]
\begin{center}
\includegraphics[width=85mm,height=75mm,scale=1.5]{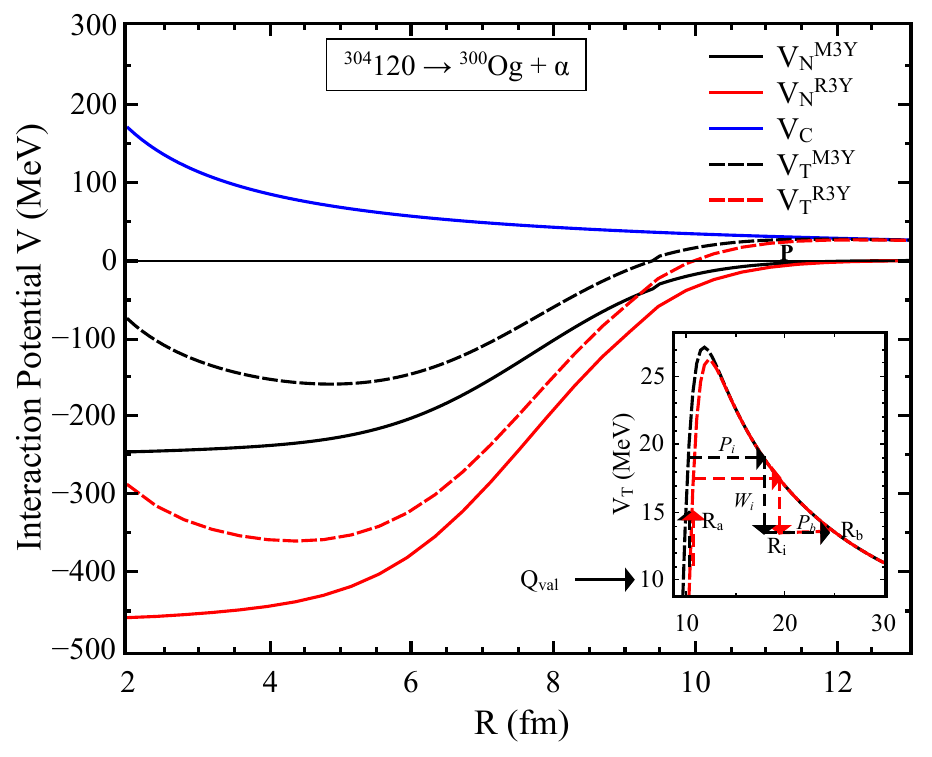}
\vspace{0.05cm}
\caption{\label{fig1} (Color online) The total ($V_{T}$) nucleus-nucleus interaction potential  (in MeV) and its components - nuclear and Coulomb potentials - are presented as a function of the radial separation $R$ (in fm) for the M3Y (NL3) and R3Y NN potentials, using the alpha decay $^{304}$120 $\rightarrow$ $^{300}$Og + $\alpha$ as a representative case. The inset provides a magnified view of the barrier height and its position corresponding to the total interaction potentials.
}
\end{center}
\end{figure}

\noindent
Nuclear physicists aim to accurately predict nuclear masses across the nuclear chart, as nuclear binding energy is critical to identifying islands of stability and magic numbers. Crucial to this effort is finding convergent solutions for the ground and excited states in the superheavy region. This involves self-consistently solving relativistic mean-field equations with varying initial deformations ($\beta_0$). The study uses 20  mesh points for Gauss-Hermite and 24 for Gauss-Laguerre integration,  employing the NL3 parameter set. The research examines the relativistic mean-field model's applicability for predicting superheavy nuclei properties and explores the effect of interaction potentials on their structural properties. In the present work, there is a study about the preformation probability and the decay properties of the five different $\alpha$-decay chains, namely, $^{296}120$ $\rightarrow$ $^{256}$Fm, $^{298}120$ $\rightarrow$ $^{258}$Fm, $^{300}120$ $\rightarrow$ $^{260}$Fm, $^{302}120$ $\rightarrow$ $^{262}$Fm, and $^{304}120$ $\rightarrow$ $^{264}$Fm. The Fig. \ref{fig1} illustrates how the nuclear potential ($V_N$), Coulomb potential ($V_C$), and total potential ($V_{T}$) vary with the distance ($R$) during the alpha decay for an illustrative case of $^{304}$120. The nuclear potential is attractive and stronger for the M3Y interaction compared to the R3Y interaction, while the Coulomb potential is repulsive and dominates at larger distances. The total potential combines these effects, forming a potential barrier at intermediate distances. The barrier is higher and wider for the M3Y interaction, leading to a classical turning point ($R_2$) at smaller radii. The smaller $R_2$ and the reduced radius at which the barrier reaches its maximum reflect the more compact nature of the potential profile for the M3Y interaction, which in turn influences the tunnelling probability of the alpha particle. The depth of the potential well and the height of the barrier are crucial for understanding the decay process and predicting the alpha particle's tunnelling probability.\\  \\
The inset of Fig.~\ref{fig1} includes the potential barrier's height and width, which are critical for determining the probability of quantum tunnelling in nuclear decay. This scattering potential is defined by three distinct processes: (a) the penetrability \( P_i \) from \( R_a \) to \( R_i \), (b) the inner de-excitation probability \( W_i \) at \( R_i \), and (c) the penetrability \( P_b \) from \( R_i \) to \( R_b \). The region between \( R_a \) and \( R_i \) represents the inner portion of the potential barrier that the $\alpha$ cluster must tunnel through to initiate the decay. The penetrability \( P_i \) is determined using the WKB approximation, which provides an estimate of the quantum tunnelling probability through a potential barrier. The probability \( W_i \) is assumed to be unity at \( R_i \), following the Greiner and Scheid model \cite{Grei86}. This model postulates that once the cluster reaches the peak of the inner barrier, it possesses sufficient excitation energy to escape without being reabsorbed. This assumption simplifies the decay analysis by focusing on the more complex external barrier. Overall, the graph provides a comprehensive visualisation of the physical factors influencing nuclear decay processes. A similar figure can be obtained for all the participating nuclei under study (but not shown here for the sake of clarity). The $Q_\alpha$-values, preformation probability, and the $\alpha$ decay half-lives of these even-even nuclei are discussed in the next subsection.\\
\subsection{{\texorpdfstring{$\alpha$}{Lg}}-decay Energy}
The $\alpha$-decay energy ($Q_{\alpha}$) is a crucial parameter for understanding decay properties and calculating the half-lives $T_{\alpha}$ of a nucleus, which provides a deeper insight into their nuclear stability. The magnitude of $Q_\alpha$ is directly connected with the decay probability and, consequently, half-life. For instance, a nucleus with a half-life exceeding the average decay timescale and exhibiting specific magic numbers of protons and neutrons at shell closures indicates increased stability. The decay energy $Q_\alpha$ can be calculated using the relations, $Q_\alpha (N, Z) = BE (2, 2) + BE (N-2, Z-2) - BE (N, Z)$. Here, BE $(N, Z)$, BE $(N-2, Z-2)$ and BE $(2, 2)$ = 28.296 MeV represent the binding energies of the parent, daughter, and $\alpha$-particle, respectively. By utilising the estimated binding energies and applying the $Q_\alpha$ equation, we can determine the decay energy for each $\alpha$-decay process. Here, the $Q-$value is obtained from the binding energy data derived from the RMF formalism for NL3 parameter set. The present study explores the $\alpha$-decay of superheavy nuclei with Z = 120 to better understand their stability and existence. The $Q_\alpha$ values are computed for various isotopes appearing in the decay chains of different isotopes of $Z=120$ nuclei, including $^{256,258,260}$Fm, $^{260,262,264}$No, $^{264,266,268}$Rf, $^{268,270,272}$Sg, $^{272,274,276}$Hs, $^{276,278,280}$Ds, $^{280,282,284}$Cn, $^{286,288}$Fl, $^{290,292}$Lv, $^{294}$Og, for which experimental Q-values \cite{wang12} are also available. Additionally, we extend this investigation to several superheavy elements (SHEs) such as $^{262,264}$Fm, $^{266,268}$No, $^{270,272}$Rf, $^{268,270,272}$Sg, $^{278,280}$Hs, $^{282,284}$Ds, $^{286,288}$Cn, $^{284,290,292}$Fl, $^{288,294,296}$Lv, $^{292,296,298,300}$Og, and $^{296-304}$120, where experimental Q-values are not available. \\
\begin{figure*}
\begin{center}
\includegraphics[width=160mm,height=95mm,scale=1.5]{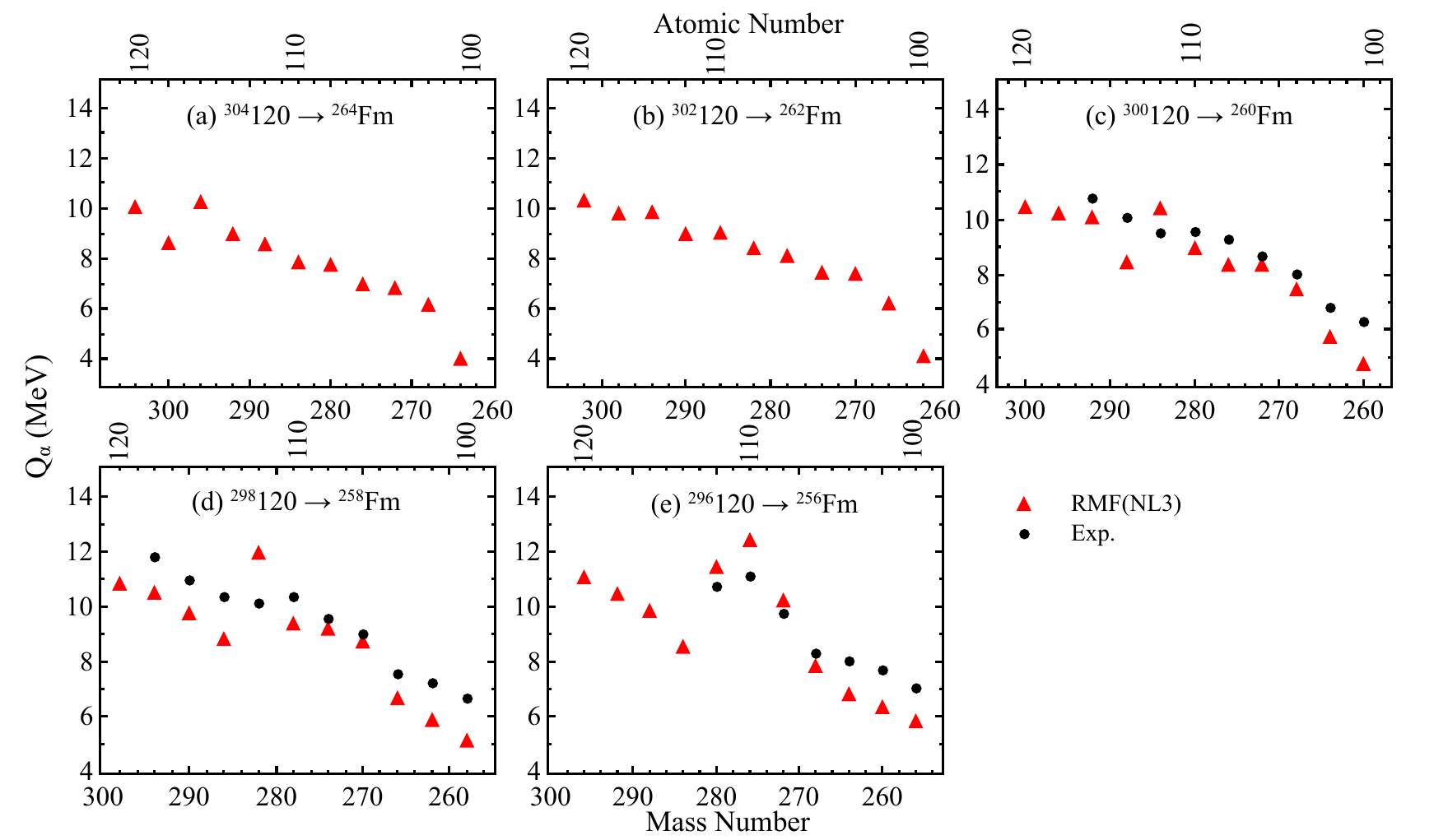}
\caption{\label{fig2} (Color online) The $\alpha$-decay (Q$_{\alpha}$) energies for five different Z = 120 chains, namely, $^{296}120$ $\rightarrow$ $^{256}$Fm, $^{298}120$ $\rightarrow$ $^{258}$Fm, $^{300}120$ $\rightarrow$ $^{260}$Fm, $^{302}120$ $\rightarrow$ $^{262}$Fm, and $^{304}120$ $\rightarrow$ $^{264}$Fm, calculated using RMF (NL3), are given along with the available experimental data \cite{wang12}. See text for details.}
\end{center}
\end{figure*}
In Fig.~\ref{fig2}, the $Q_\alpha$ values obtained from RMF (NL3) calculations are compared with available experimental data~\cite{wang12}. For most isotopes, the RMF results exhibit a close agreement with experimental values, though slight deviations appear in heavier systems. These differences mainly arise from the microscopic treatment of nuclear structure within the RMF approach, which self-consistently incorporates deformation, pairing correlations, spin--orbit coupling, and mean-field potentials.  The deviations observed in heavier nuclei reflect the sensitivity of decay energy to structural effects and underline the importance of using self-consistent models that can capture these details. The NL3 parameter set used in the present calculation was calibrated on a relatively small dataset of nuclei near the valley of $\beta$ stability, including doubly magic systems like $^{208}$Pb, and is known to provide a robust description of nuclear structure across isotopic chains. The agreement between RMF predictions and experimental data in this region validates the reliability of the approach and supports its application to the superheavy domain, where experimental information remains scarce.  Furthermore, the calculated quadrupole deformation parameters $\beta_2$ play a decisive role in shaping the ground-state configurations of superheavy nuclei. Our earlier investigations~\cite{Bisw21,Jain22} show that within the NL3 framework, nuclei in the $Z=100$--110 region exhibit decreasing $\beta_2$ values with increasing mass number, while a noticeable rise in $\beta_2$ appears for $Z=112$--120, signifying a transition toward a highly deformed prolate or even superdeformed configuration. Such transitions are essential for interpreting the structural evolution of these nuclei and could be further clarified by examining additional shape degrees of freedom, such as octupole and hexadecapole deformations~\cite{Bisw21,ahma12,frdm16,rama01,lu16}. The standard deviation in the calculated $Q_\alpha$ values for RMF (NL3) relative to experiment is found to be 0.94, indicating good consistency between theory and data. These results emphasize that the RMF model, with its self-consistent microscopic treatment, provides a reliable framework for exploring the structure and decay properties of nuclei in the superheavy region. The next subsection focuses on how the decay energy influences the preformation probability and its connection with underlying shell effects.
\subsection{Preformation Probability}
An $\alpha$-decay, a key mode of decay in heavy and superheavy nuclei, serves as a crucial probe for studying unstable, neutron-deficient nuclei and identifying newly synthesised superheavy elements. A major challenge in aligning theoretical $\alpha$-decay half-lives with experimental data is accurately calculating the $\alpha$-particle preformation factor, which depends on the nuclear shell structure and valence nucleons (holes). Recent studies \cite{josh22} highlight a linear relationship between the preformation factor and the product of valence protons and neutrons ($N_pN_n$) near shell closures ($Z$ = 82, $N$ = 126). However, existing formulas yield varied results across nuclide regions and face challenges in addressing nuclei between major shells or those with uncertain magic numbers in superheavy regions. This underscores the need for a new, comprehensive method to globally describe $\alpha$-particle preformation factors.\\
\begin{figure}[h]
\begin{center}
\includegraphics[width=85mm,height=60mm,scale=1.5]{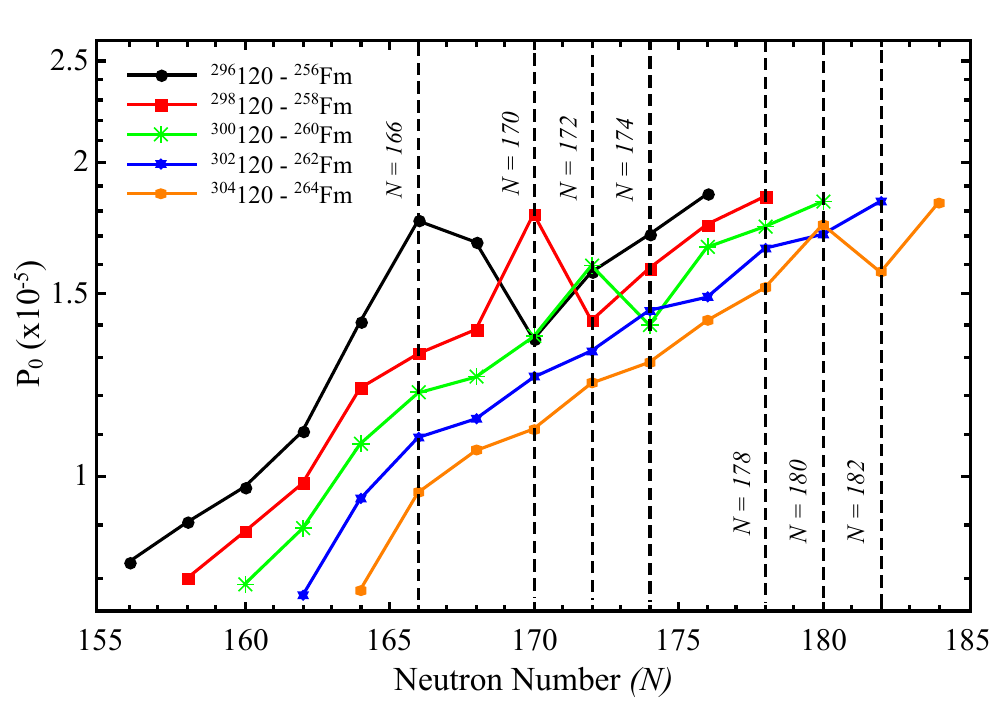}
\vspace{0.05cm}
\caption{\label{fig3} (Color online) The variation of  preformation probability w.r.t. the neutron number for five isotopes of Z = 120 chains, namely, $^{296}120$ $\rightarrow$ $^{256}$Fm, $^{298}120$ $\rightarrow$ $^{258}$Fm, $^{300}120$ $\rightarrow$ $^{260}$Fm, $^{302}120$ $\rightarrow$ $^{262}$Fm, and $^{304}120$ $\rightarrow$ $^{264}$Fm.}
\end{center}
\end{figure}
In our recent work \cite{maje23}, we introduced an analytic formula for calculating $\alpha$-particle preformation factors based on $\alpha$ decay energy $Q_\alpha$, utilising empirical formulas for $\alpha$-decay half-lives for the first time. Fig. \ref{fig3} illustrates the variations of $\alpha$-particle preformation probability $P_0$ from Eq. \ref{18} against neutron numbers $N$ for five isotopes of Z = 120 chains, namely, $^{296}120$ $\rightarrow$ $^{256}$Fm, $^{298}120$ $\rightarrow$ $^{258}$Fm, $^{300}120$ $\rightarrow$ $^{260}$Fm, $^{302}120$ $\rightarrow$ $^{262}$Fm, and $^{304}120$ $\rightarrow$ $^{264}$Fm. Each curve corresponds to a specific parent nucleus: \(^{296}\)120 (black line with circles), \(^{298}\)120 (red squares), \(^{300}\)120 (green asterisks), \(^{302}\)120 (blue stars), and \(^{304}\)120 (orange pentagons). A general increasing trend in \(P_0\) is observed for all isotopic chains as neutron number increases, which suggests that $\alpha$-cluster preformation becomes more favourable in neutron-rich isotopes, likely due to increasing binding energy and nuclear deformation effects that promote clustering. This enhancement may be attributed to reduced pairing energy and increased collectivity in heavier isotopes. However, the increase is not entirely linear, with noticeable fluctuations reflecting the influence of nuclear shell effects on stability. The graph also highlights potential magic numbers and shell closures, which manifest as kinks or plateaus in the \(P_0\) curves. \\
Focusing on individual curves, the black line corresponding to \(^{296}\)120 exhibits an early rise in \(P_0\) with a clear peak around \(N = 166\), which is associated with a local deformation-driven shell closure or enhanced pairing interaction favouring $\alpha$-cluster formation. However, this is followed by a noticeable dip at \(N = 170\), indicating a possible shell or subshell closure that impedes clustering. A similar behaviour is observed in the red line representing \(^{298}\)120, where \(P_0\) also peaks sharply at \(N = 170\), reinforcing the idea of structural effects influencing clustering. The green curve for \(^{300}\)120 shows a relatively smoother rise with a moderate peak at \(N = 172\), followed by fluctuations likely due to subtle changes in shell structure or shape coexistence phenomena in this region. The dip at $N$=174 likely arises due to shell effects that stabilise the nucleus and hinder clustering. The blue curve (\(^{302}\)120) displays a steady increase, peaking around \(N = 178\), suggesting that neutron-rich configurations in this isotope favour $\alpha$-cluster formation with minimal hindrance from shell closures. Meanwhile, the orange curve for \(^{304}\)120 maintains the lowest \(P_0\) values at lower \(N\), but eventually rises to compete with others near \(N = 180\), pointing to a gradual enhancement in clustering in the heaviest isotopes. The dip at \(N = 182\) in the orange curve (\(^{304}\)120) likely indicates the influence of a predicted neutron shell closure, which suppresses $\alpha$-cluster preformation due to increased nuclear stability. The prominent peak at \(N = 166\), particularly visible in the \(^{296}\)120 line, likely reflects a configuration where both pairing and deformation favour clustering. The peak at \(N = 170\) in the \(^{298}\)120 nucleus, followed by a dip at \(N = 172\), suggests a transition from a deformed to a more spherical shape or the onset of a shell gap that inhibits clustering.\\
Overall, the evolution of \(P_0\) with neutron number across the different Z=120 isotopes highlights the interplay between shell closures (such as at \(N =\) 172, and possibly 184), nuclear deformation, and pairing correlations. Peaks in \(P_0\) often correspond to regions where the structural features of the nucleus, either due to deformation or partial shell gaps, promote $\alpha$-cluster formation, while dips suggest stronger shell effects suppressing the same. These insights are crucial for understanding cluster radioactivity and nuclear stability in the superheavy domain, especially in the context of the elusive ``island of stability". Moreover, the $\alpha$-decay energies and half-life values are used to identify the stability of the heavy and superheavy nuclei, which are discussed in the next subsection.
\begin{figure*}
\begin{center}
\includegraphics[width=160mm,height=100mm,scale=1.5]{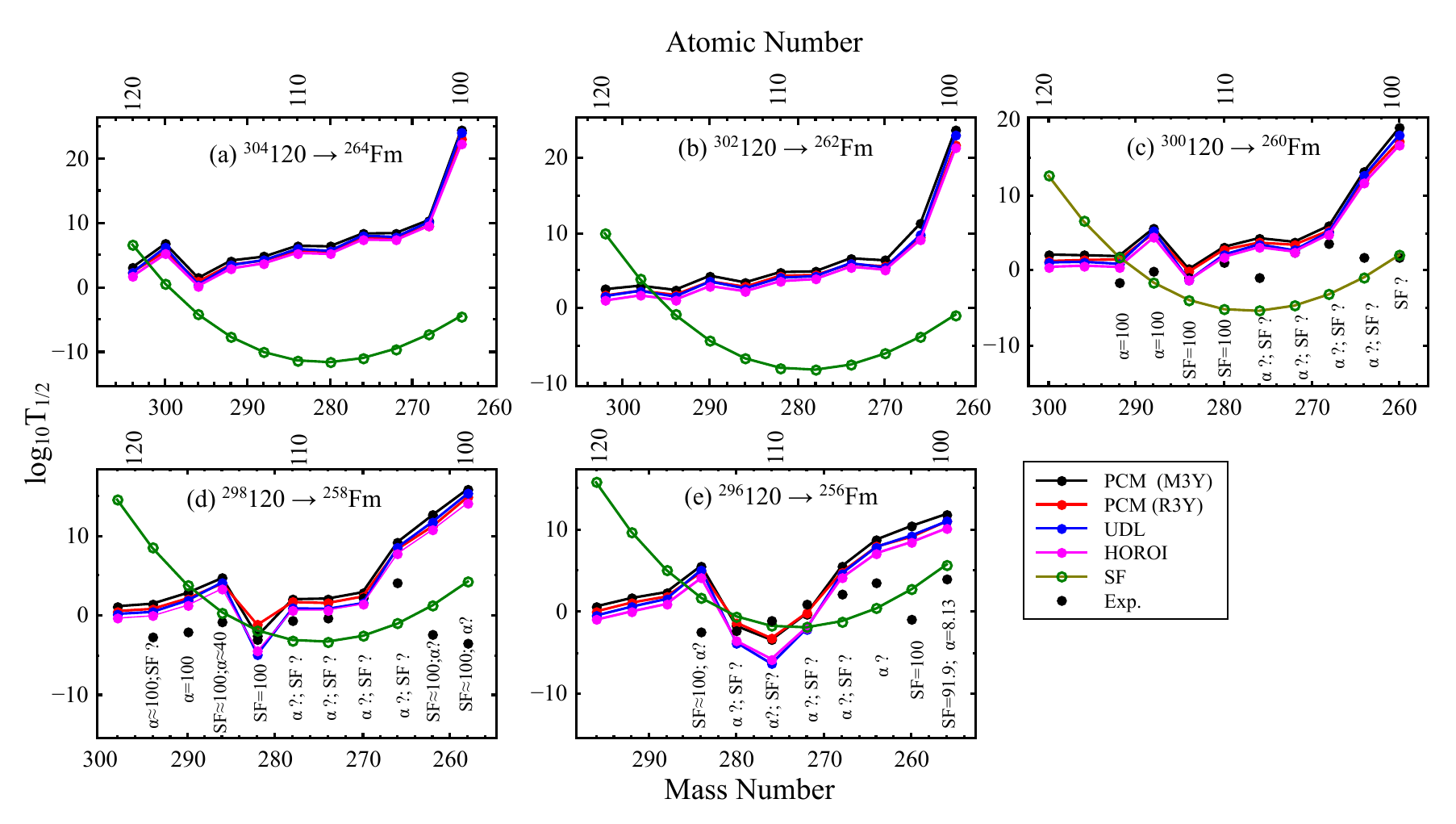}
\caption{\label{fig4} (Color online) The $\alpha$-decay half-life ($T_{1/2}^{\alpha}$) for five different Z = 120, namely, $^{296}120$ $\rightarrow$ $^{256}$Fm, $^{298}120$ $\rightarrow$ $^{258}$Fm, $^{300}120$ $\rightarrow$ $^{260}$Fm, $^{302}120$ $\rightarrow$ $^{262}$Fm, and $^{304}120$ $\rightarrow$ $^{264}$Fm, calculated using R3Y and M3Y potential are given along with the UDL \cite{isma24}, HOROI \cite{isma24} formulae, and available experimental data \cite{Audi17,audi21}. SF denotes spontaneous fission.  The decay modes and their intensities in percentages are also given along with the experimental data. A question mark (?) alone indicates a decay mode which is energetically allowed but has not yet been observed experimentally. See text for details.}
\end{center}
\end{figure*}
\subsection{{\texorpdfstring{$\alpha$}{Lg}}-decay Half-life}
The study of superheavy nuclei provides critical insights into their stability and decay modes, enhancing our understanding of their potential existence in nature. Building on these initial results, subsequent researchers have employed several methodologies to enhance the accuracy and reliability of estimating $\alpha$-decay half-lives. These methodologies also encompass analytical methods \cite{Roye00,Roye10}, semi-empirical techniques \cite{Manj17}, and empirical models \cite{Akra18b}. In this study, we conducted detailed calculations of the $\alpha$-decay and spontaneous fission half-lives for the superheavy nuclei with atomic numbers 120, spanning the isotopes from $^{296-304}120$. The main aim is to determine the predominant decay mode of each nucleus from $\alpha$ decay and spontaneous fission. Using theoretical frameworks of the Preformed cluster-decay model (PCM) and the densities obtained with the relativistic mean-field (RMF) approach for the NL3 parameter set, the present work has successfully calculated the half-lives of $\alpha$-decay and spontaneous fission (SF) of various isotopes of $Z$ = 120 nuclei.  The neck-length parameter of the PCM is fixed at $\Delta R$ = 0.54 fm for M3Y and $\Delta R$ = 1.0 fm for R3Y, which is suitable for alpha decays in terms of their respective barrier properties \cite{josh22}. We used both theoretical models and experimental data to make our analysis accurate and reliable. Each approach has aimed to modify and improve existing formulas to better correlate experimental data with the estimated $\alpha$-decay half-lives. In the present study, we have adopted a set of two recently formulated semi-empirical formulas, namely, UDL \cite{qi09} and HOROI \cite{horo04} to estimate the $\alpha$-decay half-lives of specific superheavy nuclei (SHN) under investigation. These equations are tailored to improve predictive accuracy by incorporating recent experimental results and refining the parameters influencing $\alpha$-decay. In addition, the reliable semi-empirical formula from Xu \textit{et al.} \cite{xu08} ensured accurate predictions, enabling a detailed comparison of SF as competing decay modes.\\
Fig. \ref{fig4} compares the $\alpha$-decay and spontaneous fission half-lives of $^{296-304}120$ and its $\alpha$-decay chains. The results of our calculations, conducted within the specified theoretical frameworks, are presented in Fig. \ref{fig4}(a-e) for the element with $^{296,298,300,302,304}{120}$. These results offer valuable insights into the decay characteristics of this superheavy element. It is particularly noteworthy that the $\alpha$-decay half-lives obtained using the Universal Decay Law (UDL) and HOROI formulas agree with our theoretical results based on the Relativistic Mean Field (RMF) theory for $^{296,298,300}{120}$ decay chains. The experimental decay half-lives, along with the corresponding decay modes and their intensities (in percentage), are taken from the compilation by \cite{Audi17,audi21}. The nuclei considered in our study exhibit a mixture of decay channels, primarily spontaneous fission (SF) and $\alpha$-decay, as shown in Fig. \ref{fig4}. An important aspect of the data interpretation involves distinguishing between different notations used in the decay mode information: a question mark (?) indicates that a decay mode is energetically allowed but has not yet been experimentally observed, while '=?' denotes that the decay has been observed but its intensity remains unquantified. This convention ensures a consistent and accurate representation of the experimental decay data in our analysis. Also, the present work and previous study \cite{Bisw21} show that the $Q$-values and $\alpha$-decay half-lives derived using the NL3 parameter set within the RMF framework provide a reliable fit with the experimental data \cite{Audi17,audi21} for known superheavy nuclei, wherever available. This remarkable consistency highlights the robustness and reliability of our theoretical framework, emphasising its capacity to produce accurate and reliable results under varying conditions and datasets. The alignment with experimental results highlights the efficacy of the NL3 parameter set in accurately predicting decay properties. These encouraging results make it both exciting and essential to explore the unexplored $^{302,304}{120}$ nuclei and other unknown isotopes. This extension will utilise the RMF theory in conjunction with two semi-empirical formulas, UDL and HOROI, to calculate the half-lives' decay. Exploring this unknown area could provide important insights into the stability and decay of superheavy nuclei, helping us understand the heaviest elements better. By expanding our study to include these different theoretical and empirical approaches, we aim to comprehensively and accurately predict the decay characteristics for superheavy nuclei with Z = 120 and their decay products. This work is pivotal for theoretical nuclear physics and experimental efforts aimed at synthesising and characterising new superheavy elements.\\
Fig. \ref{fig4} presents a graphical representation of $\log_{10}T_{1/2}$ plotted against the mass number of the parent nuclei, enabling a comparative analysis of the calculated $\alpha$-decay and spontaneous fission (SF) half-lives for isotopes $^{296-304}{120}$ and their respective $\alpha$-decay products. Each layer illustrates the relationship between these decay modes for different isotopic series, providing insights into their decay behaviours as follows:
\begin{enumerate}
\item Fig. \ref{fig4}(a) depicts that the isotope $^{304}{120}$ has $\alpha$-decay half-life shorter than their corresponding SF half-life. This indicates that $^{304}{120}$ is more likely to undergo $\alpha$-decay rather than fission, forming one sequential $\alpha$-decay chain from this isotope. This survival against fission highlights the relative stability of $^{304}{120}$ due to its preference for $\alpha$-decay.
\item Fig. \ref{fig4}(b) shows that the isotopes $^{302}{120}$ and $^{298}{118}$ also exhibit $\alpha$-decay half-lives shorter than their SF half-lives. Consequently, $^{302}{120}$ survives fission and forms two $\alpha$-decay chains. This trend demonstrates a consistent preference for $\alpha$-decay over SF in these isotopes, further emphasising their decay stability through $\alpha$-emission.
\item Fig. \ref{fig4}(c) illustrates the isotopes $^{300}{120}$, $^{296}{118}$, and $^{292}{116}$, all of which have $\alpha$-decay half-lives less than their corresponding SF half-lives. In particular, the isotope $^{300}{120}$ shows a strong tendency towards $\alpha$-decay, resulting in three sequential $\alpha$-decay chains. This behaviour underscores the dominance of $\alpha$-decay as the primary decay mode for these isotopes.
\item Fig. \ref{fig4}(d) highlights the isotopes $^{298}{120}$, $^{294}{118}$, and $^{290}{116}$. Again, the $\alpha$-decay half-lives are shorter than the SF half-lives, indicating that $^{298}{120}$ can undergo three $\alpha$-decay chains before any significant fission occurs. On the other hand, $^{282}{112}$ also have a half-life smaller than SF while using the PCM (M3Y), UDL and HOROI formulae. This pattern of $\alpha$-decay dominance persists, reinforcing the stability of these isotopes against fission.
\item Fig. \ref{fig4}(e) shows the isotopes $^{296}{120}$, $^{292}{118}$, $^{288}{116}$, and $^{280}{112}$. The $\alpha$-decay half-lives for these isotopes are shorter than their corresponding SF half-lives, suggesting a strong preference for $\alpha$-decay. Specifically, $^{296}{120}$ demonstrates significant stability by undergoing 
{five} $\alpha$-decay chains. 
{On the other hand, $^{276}{110}$ also have a half-life smaller than SF while using UDL and HOROI formulae.}  This extensive $\alpha$-decay chain formation further emphasises the prominence of $\alpha$-decay as the favoured decay mode in these superheavy nuclei.
\end{enumerate}
\begin{figure*}
\begin{center}
\includegraphics[width=160mm,height=85mm,scale=1.5]{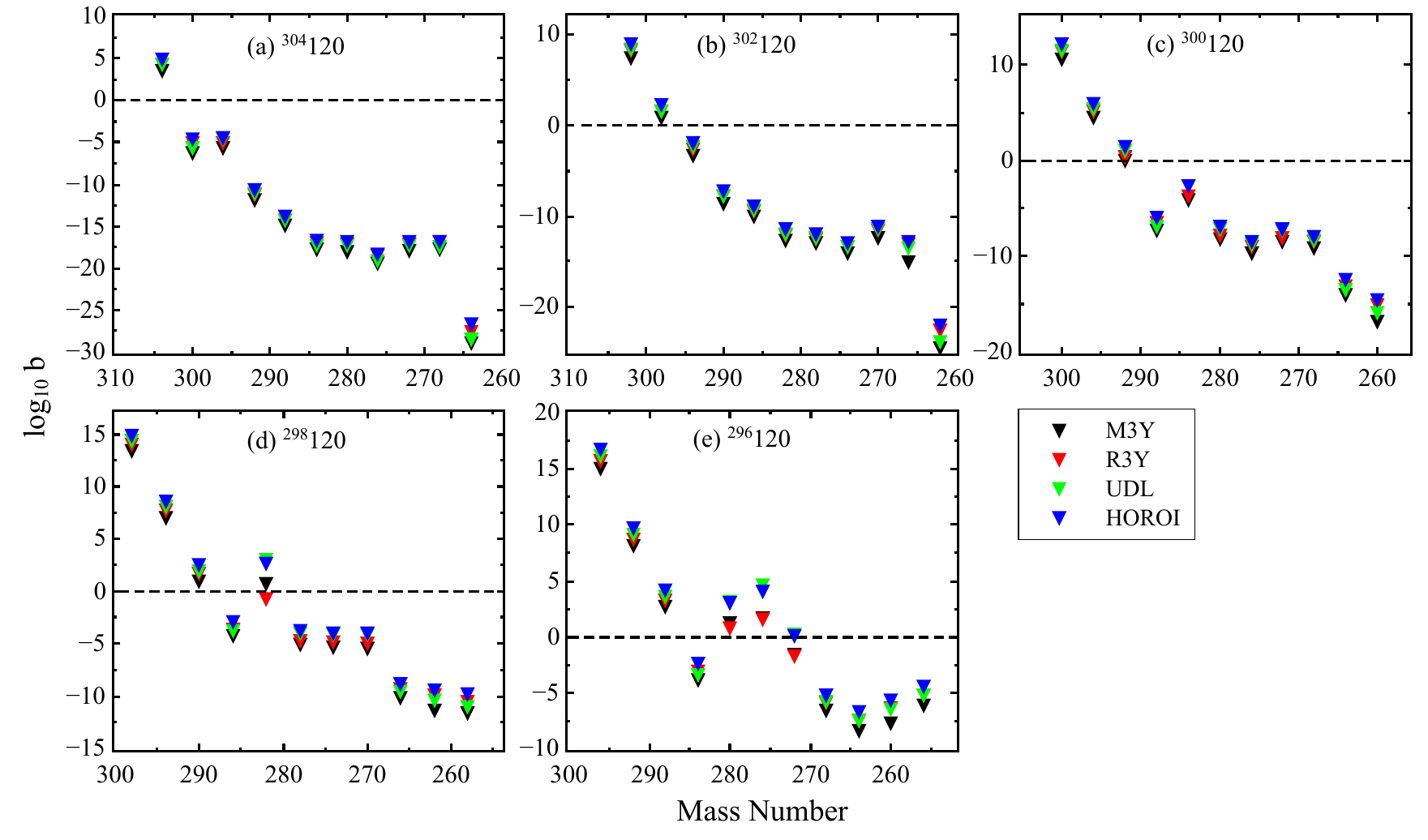}
\caption{\label{fig5} (Color online) The estimated $\log_{10}b$ for five different Z = 120 chains, namely, $^{296}120$ $\rightarrow$ $^{256}$Fm, $^{298}120$ $\rightarrow$ $^{258}$Fm, $^{300}120$ $\rightarrow$ $^{260}$Fm, $^{302}120$ $\rightarrow$ $^{262}$Fm, and $^{304}120$ $\rightarrow$ $^{264}$Fm, calculated using R3Y and M3Y potential within the PCM framework are given along with the UDL \cite{isma24}, and HOROI \cite{isma24} formulae. See text for details.}
\end{center}
\end{figure*}
The above points provide a clear and detailed comparison of the most probable decay modes of the considered nuclei. For example, consider the illustrative case of the decay $^{298}120 \rightarrow ^{258}\mathrm{Fm}$, where the difference between the measured and calculated half-life values spans nearly 5–20 orders of magnitude. This apparent discrepancy does not arise from a failure of the theoretical model, but rather from a comparison with an $\alpha$-decay mode that, although energetically allowed, has not been experimentally observed and is instead dominated by spontaneous fission (SF). According to Ref. \cite{Audi17}, the half-life of $^{258}\mathrm{Fm}$ is reported as $T_{1/2} = 370\ \mu\mathrm{s}$, with the decay mode SF $\approx 100\%$ and $\alpha$ marked as ``?”, indicating that $\alpha$-decay is energetically allowed but remains unobserved. This is consistent with the experimental findings of Hulet \textit{et al.} \cite{Hule86}, which reported that the decay of $^{258}\mathrm{Fm}$ proceeds almost exclusively via spontaneous fission, with no measurable $\alpha$-branch. Similar conditions apply to other nuclei that show significant deviations between calculated and experimental half-lives, where spontaneous fission dominates and the $\alpha$-decay branch, although theoretically allowed, lacks experimental confirmation. The calculations of $\alpha$-decay half-lives in this work assume a full $\alpha$-branching ratio (100$\%$) and are performed without any tuning or fitting to experimental half-life values, ensuring an unbiased theoretical prediction. This analysis underscores the importance of understanding the decay characteristics of superheavy nuclei, providing valuable insights into their stability and behaviour. The isotopes $^{304}{120}$, $^{302}{120}$, $^{300}{120}$, $^{298}{120}$, and $^{296}{120}$ are shown to survive fission and undergo $\alpha$-decay, highlighting their relative stability against SF and the significant role of $\alpha$-decay in their decay processes. \\
The stability of atomic nuclei is significantly influenced by their structure, particularly the presence of magic or semi-magic numbers of protons and neutrons. These magic numbers correspond to filled nuclear shells and confer additional stability to the nucleus, making it less likely to undergo decay processes such as $\alpha$ decay. The relationship between nuclear stability and these magic numbers is often represented through the examination of the half-life of the nucleus. In general, a longer half-life indicates a more stable nucleus. When we look at the logarithm of the half-life ($\log_{10}T_{1/2}$), peaks in this value can signal points of enhanced stability, which are typically associated with these magic numbers. By examining the half-lives presented in the Fig. \ref{fig4}, it becomes evident that the $\alpha$-decay chains of isotopes with the atomic number 120 (specifically $^{296-304}{120}$) exhibit distinct peaks in their stability at specific proton ($Z_p$) and neutron ($N_p$) numbers. For the isotopes $^{304}{120}$, the observed maxima in the half-life data occur at the proton and neutron number pairs ($Z_p$, $N_p$) = (118, 182), ($Z_p$, $N_p$) = (106, 168), (114, 176) for $^{302}{120}$, ($Z_p$, $N_p$) = (108, 168), (114, 174), for $^{300}{120}$, ($Z_p$, $N_p$) = (114, 172) for $^{298}{120}$, and ($Z_p$, $N_p$) = (114, 170), for $^{296}{120}$, respectively. This pattern suggests that these specific combinations of proton and neutron numbers confer enhanced stability to the nuclei, likely due to magic or semi-magic numbers \cite{Bisw21,Jain22,isma24,ahma12,adam20,prat21,jain24}. In other words, these isotopes have nucleon configurations that create a more stable nuclear structure, reducing the likelihood of $\alpha$-decay and thus extending their half-lives. For more details, we will discuss the branching ratio in the next subsection.
\subsection{Branching Ratio}
For this purpose, the branching ratio $b$ of $\alpha$ decay relative to spontaneous fission is defined as \cite{nagi20}
\begin{equation}
    b=\frac{\lambda_\alpha}{\lambda_{SF}}=\frac{T_{SF}}{T_\alpha},
\end{equation}
Here, $\lambda_\alpha$ and $T_\alpha$ represent the $\alpha$-decay constant and half-life, respectively, while $\lambda_{SF}$ and $T_{SF}$ represent the spontaneous fission (SF) decay constant and half-life, respectively. A value of $log_{10}b > 0$ indicates that $\alpha$-decay is dominant, suggesting that the particular superheavy nucleus (SHN) will survive fission and decay through $\alpha$-decay chains. Conversely, an SHN with $log_{10}b < 0$ will not survive fission. Using the predicted $\alpha$-decay half-lives from the present study and semi-empirical formulae (UDL, Horoi), as well as the data for spontaneous fission (SF) half-life, calculated using binding energies obtained through RMF with the NL3 parameter set, Figs. \ref{fig4} and \ref{fig5} were compiled to display the logarithmic half-lives and logarithmic branching ratios.
It is found that $\alpha$-decay is the dominant mode, as shown by the fact that $log_{10}b_{M3Y, R3Y} \le 0$ for 40 out of 55 superheavy nuclei (SHN). Both UDL and HOROI yield the same result, with $log_{10}b_{UDL, HOROI} \le 0$ for 39 of these SHN and differing results for 3 nuclei ($^{272}$108, $^{282}$112, and $^{292}$116). Therefore, it is concluded that the remaining 15 SHN in the Z=120 alpha decay chains will survive fission. The SHN that will survive fission are $^{276}$110, $^{280}$112, $^{288-290}$116, $^{292-298}$118, and $^{296-304}$120 according to the results obtained with the R3Y and M3Y nucleon-nucleon potentials. This analysis underscores the critical role of magic numbers in nuclear stability and the behaviour of radioactive decay processes.
\section*{Summary}
\label{result}  \noindent
A comprehensive and systematic investigation has been carried out to determine the $\alpha$-decay and spontaneous fission (SF) half-lives of superheavy nuclei (SHN) with atomic number $Z = 120$ in the mass range 256 $\leq$ A $\leq$ 304. The $\alpha$-decay half-lives ($\log_{10}T_{1/2}$) were evaluated using several theoretical approaches, including the double-folding model (DFM) with the phenomenological M3Y and the relativistic R3Y nucleon–nucleon (NN) interactions, the universal decay law (UDL), and HOROI’s scaling law. By comparing the SF half-lives with the corresponding $\alpha$-decay half-lives, the decay mode dominance across this isotopic chain has been analyzed. The results indicate that isotopes with even mass numbers $A$ = 296-304 exhibit enhanced resistance to SF, enabling the prediction of extended $\alpha$-decay chains for these nuclei. The probability of $\alpha$-cluster penetration was calculated using the semiclassical WKB approximation. The study highlights the major influence of the $\alpha$-particle preformation factor $(P_0)$ and decay energy $(Q_\alpha)$ on the underlying shell structure. The calculated $Q_\alpha$ values show strong agreement with available experimental data and assist in identifying potential new isotopes. A clear correlation is observed between the variation of $\log_{10}T_{1/2}$ and the mass numbers of parent nuclei in each $\alpha$-decay chain, with pronounced maxima near magic or semi-magic nucleon numbers. Earlier investigations \cite{Bisw21,Rutz97} identified $N = 182$ and $N = 184$ as possible magic numbers for $Z = 120$ isotopes based on structural indicators such as pairing gaps, separation energies, and shell corrections within Skyrme-Hartree-Fock and relativistic mean-field models. The present analysis shows that the R3Y interaction with the NL3 parameter set yields superior agreement across different half-life estimates. The calculated results suggest that neutron numbers $N = 166, 170, 172, 174, 176, 178, 180$, and $182$ behave as magic or quasi-magic, consistent with earlier predictions. Unpaired nucleons are also found to reduce $\alpha$-particle preformation. Overall, this study provides essential theoretical guidance for future experimental synthesis and identification of the $Z = 120$ element, offering deeper insight into the decay properties and structural trends of superheavy nuclei.
\section*{Acknowledgments}
\noindent
This work is partially supported by the Science and Engineering Research Board (SERB), File No. CRG/2021/001229, Ramanujan Fellowship File No. RJF/2022/000140, and the Hungarian National Research, Development and Innovation Office (NKFIH K134197).

\end{document}